\documentclass[pra,showpacs,preprint]{revtex4}
\usepackage{bm}
\usepackage{amsmath}
\usepackage{amssymb}
\usepackage{epsfig}

\def\be{\begin{equation}}
\def\en#1{\label{#1}\end{equation}}
\def\d{\dagger}
\def\bar#1{\overline #1}

\newcommand{\rd}{\mathrm{d}}
\newcommand{\bx}{{\bf x}}
\newcommand{\bk}{{\bf k}}

\newcommand{\bd}{{\bf d}}
\newcommand{\bQ}{{\bf Q}}
\newcommand{\bK}{{\bf K}}

\begin{document}
\title{State diagram and the phase transition of  $p$-bosons in a square  bi-partite  optical
lattice }
\author{ V. S. Shchesnovich }
\affiliation{ Centro de Ci\^encias Naturais e Humanas, Universidade Federal do ABC,
Santo Andr\'e,  SP, 09210-170 Brazil}

\begin{abstract}

It is shown that, in a reasonable approximation,  the quantum state   of $p$-bosons
in a bi-partite square two-dimensional optical lattice is governed by the nonlinear
boson model describing tunneling  of \textit{boson pairs}   between two orthogonal
degenerate quasi momenta on the edge of the first Brillouin zone. The interplay
between the lattice anisotropy  and the atomic interactions leads to the
second-order phase transition between the number-squeezed  and  coherent phase
states of the $p$-bosons. In the isotropic case of the recent experiment,
\textit{Nature Physicis} \textbf{7,} 147 (2011), the $p$-bosons are  in the
coherent phase state, where the relative global phase between the two quasi momenta
is defined only up to mod($\pi$): $\phi=\pm\pi/2$. The quantum phase diagram of the
nonlinear boson model is given.

\end{abstract}
 \maketitle

Cold atoms and Bose-Einstein condensates in  optical lattices provide a versatile
tool for exploration of the quantum phenomena of  condensed matter physics on one
hand, and, on the other hand, a way  for creation of novel types of order in cold
atomic gases \cite{OPLAT}. Two remarkable recent achievements in this direction are
the experimentally demonstrated novel types of atomic superfluids in the $P$-
\cite{Pboson} and $F$-bands \cite{Fboson} of the bi-partite square two-dimensional
optical lattice.  The bi-partite  optical lattice having a checkerboard set of deep
and shallow wells (i.e. made of double-wells), used in Refs. \cite{Pboson,Fboson}
has a large coherence time in the higher bands, several orders of magnitude larger
than the typical nearest-neighbor tunneling time \cite{CohTime}.  The order
parameter of these superfluids  is complex, in contrast to the conventional
Bose-Einstein condensates having real order parameter in accord with Feynman's
no-node theorem for the ground state of a system of interacting bosons
\cite{nonode,Wu}. The $p$-bosons, for instance,   are confined to the second Bloch
band for a sufficiently shallow lattice amplitude, $V_0 \lesssim 2.2 E_R$, where
$E_R$ is the recoil energy \cite{CohTime,InterRec}. In Ref. \cite{Pboson}
$V_0\approx 1.55E_R$, however, a particular experimental technique was used which
results in population of other Bloch bands. Nevertheless, the main results on the
cross-dimensional coherence are obtained  for  the parameter values where   the
second band is by far the largest populated.

The purpose of this work is to show  that, in the reasonable approximation,  the
quantum state of the $p$-bosons in the square bi-partite optical lattice is
governed  by the modified nonlinear boson model, which was already used before in
the context of cold atoms tunneling between the high-symmetry points of the
Brillouin zone \cite{SK,SKPRL,ParEff}. However, there is  an important difference:
in the $p$-boson case there is a lattice asymmetry parameter which provides for the
phase transition at the bottom of the energy spectrum, additionally to that at the
top of the spectrum, studied before in Ref.~\cite{SKPRL}. The focus is on the
quantum features of the $p$-boson superfluid, as different from  Ref. \cite{CW}
where a mean-field Gross-Pitaevskii approach was employed and the region of the
complex order parameter was found.

The nonlinear boson  model derived below follows just from two basic conditions:
the existence of  two quasi degenerate energy states coupled by the boson pair
exchange (tunneling) when the single-particle exchange is forbidden. Thus it
applies to other contexts as well (see also Ref. \cite{ParEff}). For instance, it
is equivalent to the nonlinear part of the so-called fundamental Hamiltonian (in
the Wannier basis), describing the local two-flavor collisions in the first excited
band of a two-dimensional single-well optical lattice \cite{CLM}. Moreover,   in
the case of the optical lattice consisting of the one-dimensional double-wells
\cite{ZPS} the many-body Hamiltonian can be cast as a set  of linearly-coupled
nonlinear boson models. Taking this into account, we consider the quantum features
of the derived nonlinear boson model in the most general setting and using its
natural parameters, besides analyzing the experimental setting  of Ref.
\cite{Pboson}.

\begin{figure}[htb]
\begin{center}
\epsfig{file=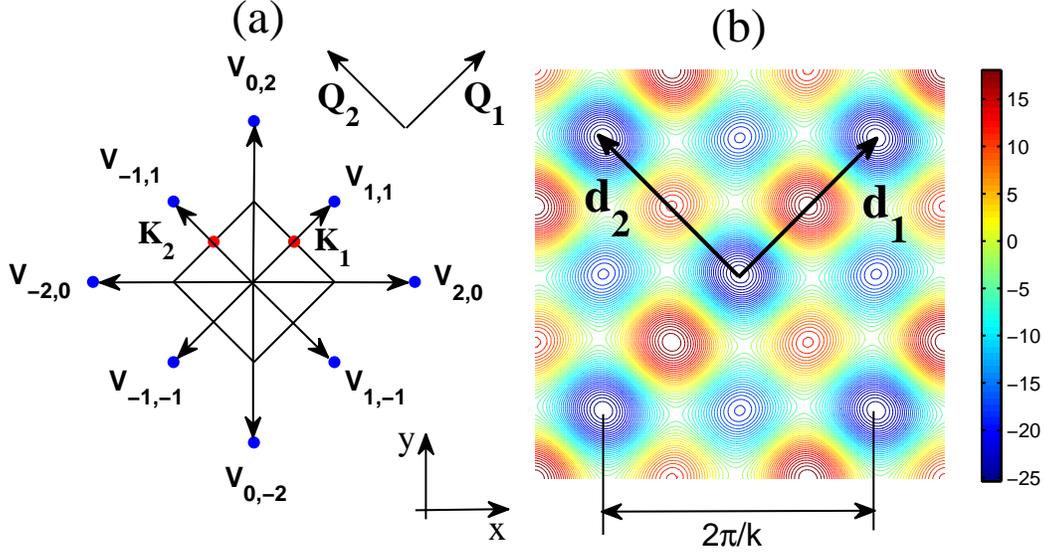,width=0.85\textwidth} \caption{(Color online) Panel (a): The
Fourier spectrum of the 2D lattice (\ref{EQ1}), where the Fourier amplitudes are
shown by the (blue) dot markers and pointed  by the reciprocal lattice vectors, the
vectors $\bQ_1=(k,k)$ and $\bQ_2=(-k,k)$ give the reciprocal lattice periods, the
points $\bK_{1,2}$, shown by two (red) dot markers, are two nonequivalent minima of
the second Bloch band. The square area is the first Brillouin zone. Panel (b): The
bi-partite square lattice in the real space, the vectors
$\bd_1=(\frac{\pi}{k},\frac{\pi}{k})$ and $\bd_2=(-\frac{\pi}{k},\frac{\pi}{k})$
form  the basis of the lattice periods.}
\label{FG1}
\end{center}
\end{figure}

 Consider the bi-partite square two-dimensional optical lattice of Ref.
\cite{Pboson}, which can be cast as follows (after dropping an inessential constant
term)
\begin{eqnarray}
V &=& -V_0\exp\left(-\frac{z^2}{a^2_z}\right)\bigl(2v_{2,0}\cos(2kx) +
2v_{0,2}\cos(2ky)
\nonumber\\
&& + 2\mathrm{Re}\{v_{1,1}e^{ik(x+y)} + v_{-1,1}e^{ik(y-x)} \}\bigr),
\label{EQ1}
\end{eqnarray}
where the experimental values of the parameters read $V_0 =  \bar{V}_0/4=1.55 E_R$
in terms of the recoil energy \mbox{$E_R = \frac{\hbar^2 k^2}{2m}$,} with  $k =
2\pi/\lambda$, $\lambda = 1064\,$nm and $a_z = 71\,\mu$m   being the oscillator
length of the transverse trap. The dimensionless Fourier amplitudes of the lattice
are
\begin{eqnarray}
&&v_{2,0} = \eta^2\epsilon\cos\alpha,\quad v_{1,1} =
\eta\epsilon\left[e^{i\theta}+e^{-i\theta}\cos\alpha \right],
\nonumber\\
&& v_{0,2} = \epsilon,\quad
v_{-1,1}=\eta\left[e^{i\theta}\cos\alpha +\epsilon^2e^{-i\theta}\right],
\label{EQ2}
\end{eqnarray}
see Fig. \ref{FG1}. The experimental parameters are $\eta\approx 0.95$ and
$\epsilon \approx 0.81$.

For $\alpha_{iso}=\arccos\epsilon\approx \pi/5$ and  arbitrary values of the other
parameters we have $v_{1,1} = v_{-1,1}$, hence, the lattice satisfies the symmetry
$V(-x,y,z) = V(x,y,z)$. For  $\alpha= \alpha_{iso}$ the  band energies with the
Bloch indices $\bK_1=(\frac{k}{2},\frac{k}{2})$ and
$\bK_2=(-\frac{k}{2},\frac{k}{2})$ (see Fig. \ref{FG1}(a)) become equal, since the
Bloch functions satisfy $\varphi_{\bK_1}(-x,y,z) = \varphi_{\bK_2}(x,y,z)$, due to
the boundary conditions and symmetry of $V(x,y,z)$. Hence, the   points $\bK_{1,2}$
are the high-symmetry points of the symmetric lattice with  (see also Ref.
\cite{SK}).

As is found in Ref. \cite{CW} the observed cross-dimensional coherence
\cite{Pboson} is the joint effect of the atomic interactions and the lattice
potential. Indeed, let us estimate the interaction energy and its characteristic
time scale in the $p$-boson experiment. The interaction energy can be estimated as
$E_{int} \sim \frac{gN}{\Omega}$, where $g = \frac{4\pi \hbar^2 a_s}{m}$ is the
interaction coefficient proportional to the $s$-wave scattering length $a_s$, $N$
is the number of atoms and $\Omega$ is the effective volume of the condensate.
Setting $\Omega =\sqrt{2\pi} a_z L^2 \lambda^2$ (the coefficient is due to ground
state of the transverse trap, see below), where $L$ is the number of the lattice
sites along each of the two directions $\bd_{1,2}$ in the plane $\bx \equiv (x,y)$,
we get $E_{int} \sim \frac{a_s N}{a_z L^2}E_R$. For ${}^{87}$Rb and other
experimental values of Ref. \cite{Pboson}, with $N\sim 10^5$ and $L \sim 10$ (the
estimated  sample size of Ref. \cite{Pboson} divided by the lattice cell size), we
get $E_{int}/E_R\sim 0.1$. Moreover, the interaction time scale, defined as
$t_{int}= \hbar/E_{int}$ is on the order of the typical experimental times, indeed,
we have $t_{int}\sim \frac{m a_z L^2 \lambda^2}{\hbar a_s N} \sim 10\,$ms (compare
with Fig. 2 of Ref. \cite{Pboson}).

Taking into account the above estimate,  one can assume that the atoms are confined
to the second Bloch band of the   lattice   and expand the boson field operator
over the band-limited  Bloch basis. The Bloch waves   are defined as
$\varphi_\bk(\bx) = \frac{1}{L}e^{i\bk\bx}u_{\bk}(\bx)$, $u_{\bk}(\bx+\bd_j)=
u_{\bk}(\bx)$, $j=1,2$, where   the periodic Bloch functions $u_{\bk}(\bx)$ are
chosen to be  normalized on the 2D lattice cell $\nu_0=|\bd_1\times\bd_2|$ of area
$\lambda^2/2$, i.e.
\[
\int\limits_{\nu_0}\rd^2\bx |u_{\bk}(\bx)|^2=1.
\]
The band-limited expansion reads
\be
\Psi(\bx,z) =\sum_{\bk\in B\!Z} b_\bk \varphi_{\bk}(\bx)\Phi_0(z),
\en{EQ3}
where the summation  is over the Bloch indices inside the first Brillouin zone $\bk
\in\{\frac{\kappa_1}{L}\bQ_1 + \frac{\kappa_2}{L}\bQ_2;\;
\kappa_j=-L/2,\ldots,L/2\}$  (see, Fig. \ref{FG1}(a)). Here $\Phi_0(z)$ is  ground
state \mbox{$\Phi_0(z) \approx \pi^{-1/4}a_z^{-1/2}e^{-z^2/(2a_z^2)}$} of the
transverse trap. Inserting this expression into the standard Bose-Hubbard
Hamiltonian for the lattice potential (\ref{EQ1}) and using the Poisson summation
formula,
\[
\sum_{\ell_1,\ell_2=1}^L e^{i\bk(\ell_1\bd_1+\ell_2\bd_2)} = L^2\delta_{\bk,\bf{0}}
\quad \mathrm{mod}(\bQ),
\]
we obtain
\begin{widetext}
\begin{eqnarray}
H &=& \sum_{\bk\in B\!Z} E(\bk)b^\d{}_\bk b_\bk + \frac{g}{2\sqrt{2\pi}
a_z}\int\rd^2\bx\left\{\left[\sum_{\bk\in B\!Z} b^\d_\bk
\varphi^*_{\bk}(\bx)\right]^2\left[\sum_{\bk\in B\!Z} b_\bk
\varphi_{\bk}(\bx)\right]^2\right\}\nonumber\\
&=&\sum_{\bk\in B\!Z} E(\bk)b^\d{}_\bk b_\bk +
\frac{g}{2\Omega}\sum_{\Delta_{\bk}=\bf{0}}
\chi(\bk_1,\bk_2|\bk_3,\bk_4)b^\d{}_{\bk_1}b^\d{}_{\bk_2}b_{\bk_3}b_{\bk_4},
\label{EQ4}
\end{eqnarray}
\end{widetext}
where $E(\bk)$ is the Bloch energy of the second band, $\Delta_{\bk}\equiv
\bk_1+\bk_2-\bk_3-\bk_4$, the condition $\Delta_\bk =\bf{0}$ is understood
mod($\bQ)$  and
\be
\chi(\bk_1,\bk_2|\bk_3,\bk_4)
=\nu_0\int\limits_{\nu_0}\rd^2\bx\,u^*_{\bk_1}u^*_{\bk_2}u_{\bk_3}u_{\bk_4}
\en{EQ5}
is  the  dimensionless  coefficient  which depends solely on the lattice geometry.

Since  the points $\bK_{1,2}$, the  energy   minima of the second band, are lying
on the edge of the Brillouin zone (Fig. \ref{FG1}(a)),   the Bloch functions
$\varphi_{\bK_{1,2}}(\bx)$ are real. Moreover  $\nabla_{\bk}E(\bk)=0$ and, hence,
$\nabla_{\bk} \varphi_{\bK_{1,2}}(\bx)=0$. As the result, the expansion over $\bk$
in Eq. (\ref{EQ4})  in some small neighborhoods about these points starts only with
the second-order term \mbox{$\propto(\bk-\bK_{1,2})^2$} \cite{FootNote1}. On the
other hand, one can verify that  the experimental width of the Bragg peaks about
the band minima $\bK_{1,2}$ is too narrow to give a significant second-order
correction, i.e.   $(\bk-\bK_{1,2})^2/k^2\sim 0.06$ (see, Fig. 3 of Ref.
\cite{Pboson}). Therefore, we can discard the spectral width of the Bragg peaks and
keep in Eq. (\ref{EQ3}) only the  two-mode expansion of the boson field operator (a
similar expansion over the two nonlinear modes  was also used in Ref. \cite{CW})
\be
\Psi(\bx,z) \approx \left[b_1 \varphi_{\bK_{1}}(\bx) +
b_2\varphi_{\bK_{2}}(\bx)\right]\Phi_0(z).
\en{EQ6}
It is important to note that, since the  summation in the nonlinear term of Eq.
(\ref{EQ4}) is conditioned by  $\Delta_\bk$=0 mod$(\bQ$), all terms with  with
either three $\bK_{1}$ and one $\bK_{2}$, or vice versa are zero (i.e. bosons
tunnel between the minima by \textit{pairs} \cite{SK}). Thus, only the following
geometric parameters are nonzero:
\begin{eqnarray}
&&\chi_{jj} = \nu_0\int\limits_{\nu_0}\rd^2\bx|u_{\bK_{j}}(\bx)|^4,\quad j=1,2,\nonumber\\
&&
\chi_{12} = \nu_0\int\limits_{\nu_0}\rd^2\bx|u_{\bK_{1}}(\bx)u_{\bK_{2}}(\bx)|^2.
\label{chis}\end{eqnarray}
As a consequence,  one obtains from Eq. (\ref{EQ4}) the two-mode Hamiltonian of the
nonlinear   boson model \cite{SK,SKPRL,ParEff} except for the term  proportional to
the population imbalance  due to the lattice asymmetry:
\begin{eqnarray}
H &=& \frac{\mathcal{E}_1-\mathcal{E}_2}{2}(n_1 - n_2) +\frac{U}{2}\bigl\{ n_1(n_1-1) +n_2(n_2-1)\nonumber\\
&&+ \Lambda(4n_1n_2 + (b^\d{\!}_1b_2)^2 + (b^\d{\!}_2b_1)^2 )\bigr\}.
\label{EQ7}
\end{eqnarray}
We have denoted $n_{j} \equiv b^\d_{j} b_{j}$. The parameters of Hamiltonian
(\ref{EQ7}) are as follows. The  energies of the two symmetric points $\bK_{1,2}$
read
\be
\mathcal{E}_{1} = E_1 + \frac{\chi_{11}}{2}\frac{gN}{\Omega},\quad \mathcal{E}_{2}
=  E_2 + \frac{\chi_{22}}{2}\frac{gN}{\Omega},
\en{EQ8-1}
where $E_{1,2} = E(\bK_{1,2})$  is the respective Bloch energy, $N= n_1 + n_2$, $U$
is  the average interaction parameter per particle,
\be
U = \frac{g}{2\Omega}(\chi_{11}+\chi_{22}),
\en{EQ8-2}
and $\Lambda$ is  a pure geometric parameter defined as
\be
\Lambda = 2\chi_{12}/(\chi_{11}+\chi_{22}).
\en{EQ8-3}
Note that at the symmetric point $\alpha=\alpha_{iso}$ we have $\sigma =0$, hence
$\mathcal{E}_1=\mathcal{E}_2$. We have just two independent parameters
$(\gamma,\Lambda)$, where $\gamma$ is defined as
\be
\gamma = \frac{\mathcal{E}_1-\mathcal{E}_2}{UN} = \frac{E_1-E_2}{UN} + \sigma,
\quad \sigma\equiv \frac{\chi_{11}-\chi_{22}}{\chi_{11}+\chi_{22}}.
\en{EQ9}
Here we note that  any 2D lattice which for some set of parameters possesses two
non-equivalent points lying on the edge of the Brillouin zone and having equal
Bloch energies   can lead, under similar conditions, to the same model Hamiltonian
(\ref{EQ7}).

\begin{figure}[htb]
\begin{center}
\epsfig{file=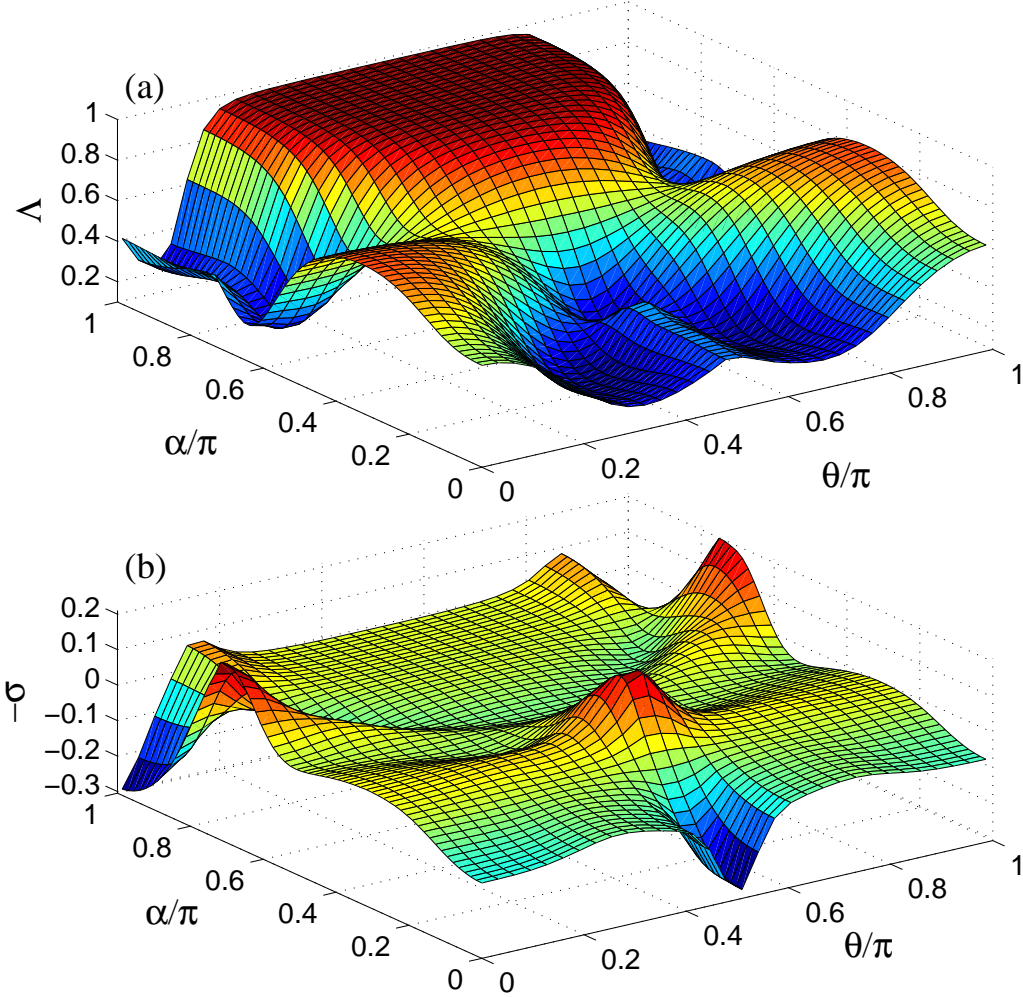,width=0.85\textwidth} \caption{(Color online) The
numerically computed lattice parameters $\Lambda$ (panel (a)) and $\sigma$
(panel(b)) for the experimental 2D lattice (\ref{EQ1}) of Ref. \cite{Pboson}. Here
$V_0=1.55E_R$, $\eta=0.95$, $\epsilon = 0.81$. (The high accuracy Fourier
pseudospectral method \cite{Pseudo} was used.) }
\label{FG2}
\end{center}
\end{figure}

\begin{figure}[htb]
\begin{center}
\epsfig{file=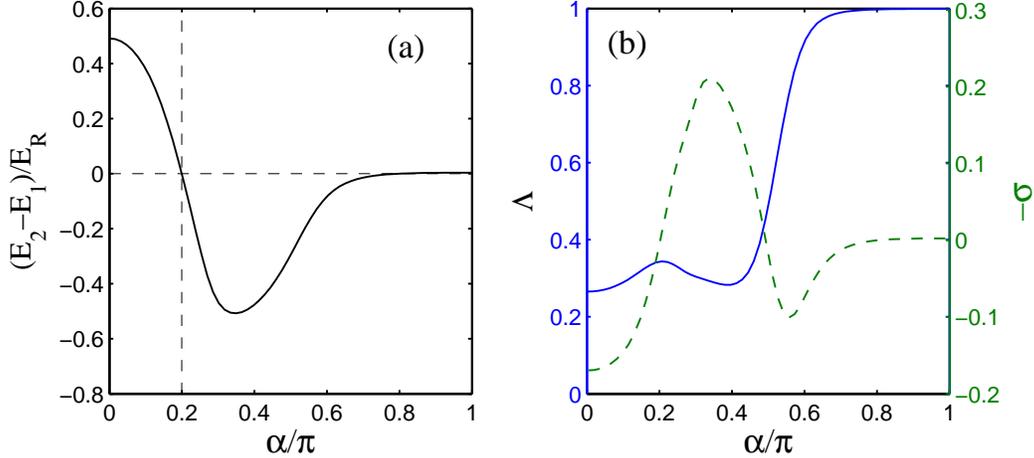,width=0.85\textwidth} \caption{(Color online)  The energy
difference, (a), and the lattice parameters $\Lambda$ (the left $y$ axis, solid
line) and $\sigma$ (the right $y$ axis, dashed line), (b). Here $\theta=0.53\pi$
and the other parameters are as in Fig. \ref{FG2}.}
\label{FG3}
\end{center}
\end{figure}

The parameters $\Lambda$ and $\sigma$, $0\le \Lambda \le 1$ and $-1\le \sigma \le
1$, are independent of the interaction strength $g$ and are functions  only of the
lattice shape. For the experimental lattice (\ref{EQ1}) their dependence on
 $\alpha$ and $\theta$ can be  determined by numerically solving the 2D
eigenvalue problem for Bloch energies, the result   is given in Fig. \ref{FG2}.
Except for the  semicircle shaped plateau, both parameters vary significantly with
variation of the lattice potential. Specifically, for the   experimental value
$\theta=0.53\pi$  the parameters $\Lambda$, $\sigma$ and the Bloch energy
difference  are given in Fig.~\ref{FG3}.

The interaction energy parameter $UN$  was already estimated above, i.e. $UN/E_R
\sim {a_s N}/(a_z L^2)\sim 0.1$ for the experimental values of Ref.~\cite{Pboson},
the bandgap is on the order of the lattice amplitude $V_0\sim E_R$, $\sigma\sim
0.2$, see Fig.~\ref{FG2}(b),  whereas the energy degeneracy is at most $\sim
0.4E_R$, see Fig.~\ref{FG3}(a). We conclude that in the experiment of
Ref.~\cite{Pboson} $\gamma$  can reach order $\sim 1$.

The ground state of the model Hamiltonian (\ref{EQ7}) for $\gamma\ne0$ can be one
of the two types of states: either  the  coherent  phase state (with a definite
relative phase between the two modes $b_{1,2}$, $\phi = \pm\pi/2$) or the atom
number squeezed (Bogoliubov) state. These two types of the ground state are
connected by the second-order quantum  phase transition on the   borderlines
$\Lambda = 1+|\gamma|$ in the plane $(\gamma,\Lambda)$ (see also Fig.~\ref{FG4}
below). There are, in fact, exactly two phase transitions. One is at the bottom of
the quantum energy spectrum and occurs for the asymmetry parameter $\gamma\ne 0$
(it corresponds to the relative phase $\pm\pi/2$, see below). The other one is at
the top of the spectrum (and corresponds to the zero relative phase). For
$\gamma=0$ the phase transition at the top of the spectrum was studied before
\cite{SKPRL,ParEff}.

Consider first the  number-squeezed states, which appear for the  large population
imbalance between the points $\bK_{1,2}$ and  have a squeezed variance of the
population  imbalance  (see also Refs. \cite{SKPRL,ParEff}). For instance, suppose
that $n_1\gg n_2$ (i.e. $n_1\approx N$) and  denote the respective class of states
by $B_1$. Following Bogoluibov's approach, one can replace $b_1 \to
\sqrt{N-n_2}e^{i\Phi}$, where $\Phi$ is an inessential random   phase, and expand
the Hamiltonian (\ref{EQ7}) in orders of $b_2$ and $b^\d_2$. Keeping the
second-order terms only we get the local quadratic Hamiltonian in the form
$H\approx\frac{UN^2}{2}\hat{H}_1$ with
\be
\hat{H}_1 =(1+\gamma) +2(2\Lambda-1 +\gamma)\frac{n_2}{N} +
\frac{\Lambda}{N}[(e^{i\Phi}b^\d_2)^2+(e^{-i\Phi}b_2)^2].
\en{EQ10}
The Hamiltonian (\ref{EQ10}) is diagonalizable by the Bogoliubov transformation
\be
b_2=e^{i\Phi}[\cosh\!\beta a- \sinh\!\beta a^\d],\; \tanh(2\beta) =
\frac{\Lambda}{2\Lambda-1 -\gamma}.
\en{EQ10A}
where $\beta$ is the squeezing parameter. We have
\be
\hat{H}_1 = 1+\gamma + \frac{\Lambda}{N}\left[\frac{2 }{\sinh(2\beta)}\left(a^\d a
+\frac12\right) -  \coth(2\beta)\right].
\en{EQ10B}
For $n_2\gg n_1$  the   number-squeezed states $B_2$ are described by similar
quadratic Hamiltonian $\hat{H}_2$ obtained by replacing $b_2$ with $b_1$ in Eqs.
(\ref{EQ10})-(\ref{EQ10A}) as well as  inverting the sign at $\gamma$.

The existence diagram  of the number-squeezed states $B_{1,2}$  is shown  in Fig.
\ref{FG4}(a), their existence is equivalent to existence of  the Bogoliubov
transformation (\ref{EQ10A}). The  states $B_{1,2}$ are  thermodynamically stable
for positive effective mass in Eq. (\ref{EQ10B}), i.e. when $ 2\Lambda-
1\mp\gamma>0$,  which condition is  satisfied only in the regions
$\Lambda>1+\gamma$ and $\Lambda>1-\gamma$, respectively for $B_1$ and $B_2$. The
thermodynamically stable $B_{1,2}$ states are shown in Fig.~\ref{FG4}(d). Note that
the number-squeezed states have undefined relative phase $\phi =
\mathrm{arg}(\langle (b^\d_1)^2b_2^2\rangle)/2$ (this is reflected also in
arbitrariness  of $\Phi$, see also the discussion of the quantum phase below).

Hamiltonian (\ref{EQ7})  also admits the phase states possessing definite values
(i.e. with small variance) of the phase   and  the    population imbalance. These
states will be called coherent. The existence  diagram of the coherent states can
be found by approximating the Hamiltonian by a quantum oscillator problem in the
Fock space \cite{SKPRL}. For $N\gg1$ the coherent states are essentially
semiclassical in the sense of Ref. \cite{Braun}. Thus, most of their properties can
be studied by replacing the boson operators by scalar amplitudes: $b_1\to
\sqrt{N/2(1+\zeta)}e^{-i\phi/2}$ and $b_2\to \sqrt{N/2(1-\zeta)}e^{i\phi/2}$ and
considering the resulting classical model (save for the factor $\frac{UN^2}{2}$)
\be
\mathcal{H}_\mathrm{cl} = \frac12 + \gamma\zeta + \frac12\bigl[\zeta^2 +
\Lambda(1-\zeta^2)(2+\cos(2\phi))\bigr].
\en{Hcl}
The stable stationary points of the classical Hamiltonian $\mathcal{H}_\mathrm{cl}$
correspond to the phase states of the quantum model. There are  two stationary
points: $(2\phi_t=0,\zeta_t=\frac{\gamma}{3\Lambda-1})$ and $(2\phi_b = \pi,\zeta_b
= -\frac{\gamma}{1-\Lambda})$ and they correspond, respectively, the coherent phase
states  at the top  ($C_0$) and at the bottom ($C_\pi$) of the quantum energy
spectrum (this is clear from their energies).

The direct  approach to study the coherent states is based on  the discrete WKB in
the Fock space, with the effective Planck constant $h = 2/N$ \cite{SKPRL}. One
first factors out the classical phase $\phi_{b,t}$ and then expands the Hamiltonian
(\ref{EQ7}) about the classical stationary point $\zeta_{b,t}$ (see also Ref.
\cite{ST}). Representing the  Fock-space ``wave function'' $\psi(\zeta) = \langle
n,N-n|\psi\rangle$ (here $n\equiv n_1$) with $\zeta = 2n/N-1$ as $\psi =
e^{i\phi\zeta/h}\psi_0(\zeta)$ and defining the canonical with $\zeta$ momentum as
$\hat{p} = -ih\partial_\zeta $ we get
\be
\langle n,N-n|H|\psi\rangle  = e^{i\phi\zeta/h}
\frac{UN^2}{2}\hat{H}_{\phi}\psi_0(\zeta),
\en{EQ11}
with a local  Hamiltonian $\hat{H}_\phi$ of a quantum oscillator (the discarded
terms start with $ \sim(\zeta-\zeta_{b,t})^3$). The Hamiltonian   about $\zeta_{b}$
(for the phase $2\phi_b=\pi$) reads
\be
\hat{H}_{\phi_b} = \frac{1-\gamma^2-\Lambda^2}{2(1-\Lambda)}
+\frac{1-\Lambda}{2}(\zeta-\zeta_b)^2+\frac{\Lambda}{4}(1-\zeta_b^2)\hat{p}^2,
\en{EQ12}
while that about the point $\zeta_t$ (for $2\phi_t=0$) can be obtained by replacing
$\Lambda$ by $3\Lambda$ in the first two terms in Eq.~(\ref{EQ12}) and inverting
the sign at $\hat{p}^2$ due to the negative effective mass  $M_{b,t}^{-1} =
-\Lambda/2(1-\zeta_{b,t}^2)\cos(2\phi_{b,t})$. The existence and stability analysis
is straightforward from this point. First of all, the coherent states $C_0$, i.e.
with the classical phase satisfying $2\phi_t = 0$, are thermodynamically unstable
due to the negative effective mass, while the states $C_\pi$ are thermodynamically
stable where they exist. The existence diagram of the coherent states   is given in
Figs. \ref{FG4}(b) and (c). Numerical simulations confirm that the Gaussian width
of the oscillator ``wave-function'' $\psi(\zeta)$ reasonably approximates the width
of the coherent states in  the Fock space.

\begin{figure}[htb]
\begin{center}
\epsfig{file=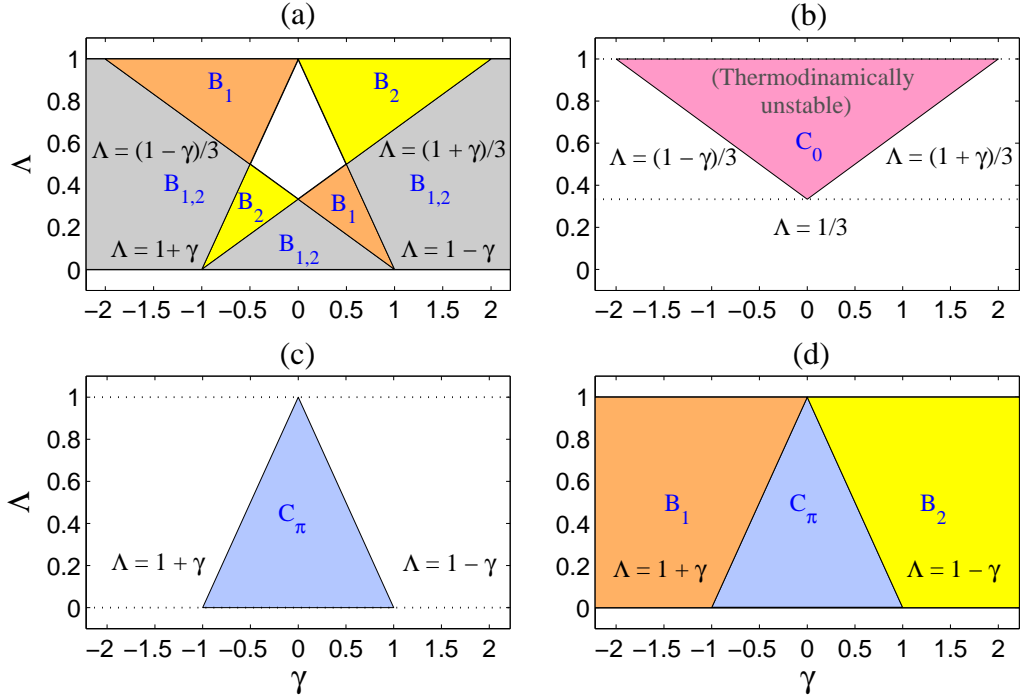,width=0.85\textwidth} \caption{(Color online)  (a) The
existence diagram  of the  number-squeezed (Bogoliubov)   states $B_{1,2}$ of
Hamiltonian (\ref{EQ7}). (b) The coherent states  of zero relative phase,
corresponding to the top of the quantum energy spectrum ($C_0$). (c) The coherent
states with the relative phase $\phi = \pm\pi/2$, corresponding to the bottom of
the quantum energy spectrum ($C_\pi$). (d) The ground state diagram of the model
Hamiltonian (\ref{EQ7}). }
\label{FG4}
\end{center}
\end{figure}

By considering the characteristic  energies (up to $\sim 1/N$) in terms of $UN^2/2$
of all the above classes of states, i.e. $E(B_{1,2}) = 1\pm\gamma$, $E(C_{\pi}) =
\frac{1-\gamma^2-\Lambda^2}{2(1-\Lambda)}$ and $E(C_0) =
\frac{1-\gamma^2-9\Lambda^2}{2(1-3\Lambda)}$, one obtains the  state diagram of the
model (\ref{EQ7}), see Fig. \ref{FG4}(d). Depending on the values of $\gamma$ and
$\Lambda$ the  ground state   is either  the coherent state $C_\pi$ or one of the
squeezed states, $B_{1} $ or $B_2$. The phase transition borderline is $\Lambda =
1+|\gamma|$. Figs.~\ref{FG4}(a) and (b) demonstrate  that a similar phase
transition occurs at the top of the energy spectrum on the border line given by
$\Lambda = (1+|\gamma|)/3$. It was the subject of Refs. \cite{SKPRL,ParEff}.

\begin{figure}[htb]
\begin{center}
\epsfig{file=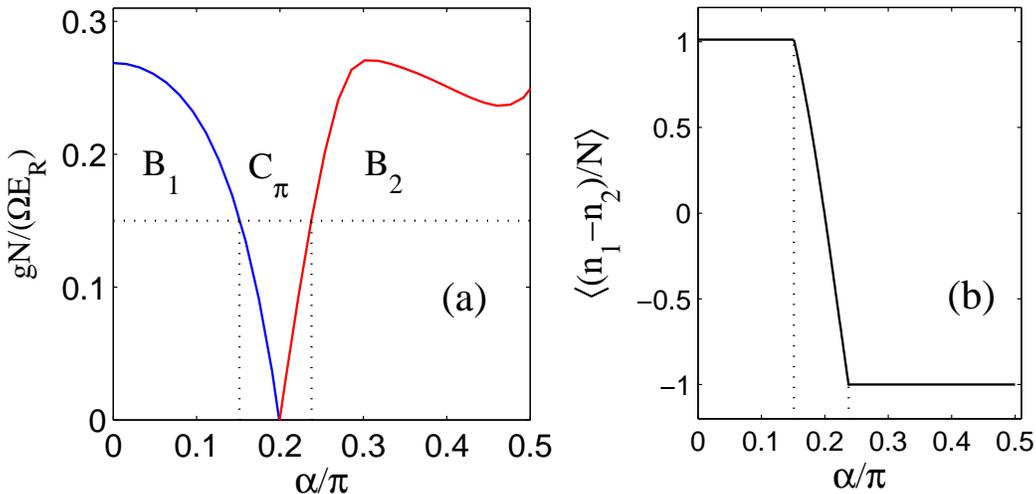,width=0.85\textwidth} \caption{(Color online)  (a) The phase
diagram  of  the model Hamiltonian (\ref{EQ7}) in terms of the effective
interaction parameter $g\frac{N}{\Omega}$ in units of the recoil energy vs. the
angle $\alpha$. Here $\theta = 0.53\pi$ and the rest of the parameters as in Fig.
\ref{FG2}. (b) The typical average relative population imbalance between the two
points $\bK_{1,2}$, which  correspond to the semiclassical $\zeta_b$ between two
values of $\alpha$  (the dotted lines in (a) and (b)). }
\label{FG5}
\end{center}
\end{figure}

Let us now consider the state diagram versus the experimental parameter $\alpha$.
To compare the result also to the mean-field diagram of Ref. \cite{CW} (see Fig. 5)
one has to identify the same interaction parameter (the product of the $g$ and the
density in Ref. \cite{CW}).  The quantity $gN/\Omega$ can serve as an analog,
though one has to remember that we have discarded the atoms of the condensate not
represented by the Bragg peaks at the two points $\bK_{1,2}$, thus the resulting
approximate  value of $gN/\Omega$ will be  smaller than the actual value and the
comparison can be only qualitative. The expressions for the borderlines $\Lambda =
1\pm\gamma$ of the state diagram Fig.~\ref{FG4}(d) can be rewritten using
Eqs.~(\ref{EQ8-1}), (\ref{EQ8-2}) and (\ref{EQ9}) as to give the interaction
parameter $gN/\Omega$. We obtain:
\begin{eqnarray}
&& g\frac{N}{\Omega} = \frac{E_2 - E_1}{\chi_{11} - \chi_{12}},\quad \gamma\le0,
\label{EQ19}
\\
&&g\frac{N}{\Omega} = \frac{E_1 - E_2}{\chi_{22} - \chi_{12}},\quad \gamma\ge0.
\label{EQ20}
\end{eqnarray}
The  results are presented in Fig. \ref{FG5}, where the energy is given in the
recoil energy units. Qualitatively we have similar diagram to that of Ref.
\cite{CW}, though  the corresponding quantitative value of the interaction
parameter $gN/\Omega$ is significantly smaller (though  the   density parameters
are not identical, as mentioned above,  the difference is still significant). We
note, however, that the values of the interaction parameter in Fig.~\ref{FG5}   do
correspond to the estimated value $gN/\Omega \sim UN \sim 0.1E_R$ which accounts,
for instance,  for  the Bragg peak formation times. This estimate was used to
validate the expansion (\ref{EQ6}) over the Bloch modes, which was then used in the
nonlinear part of the many-body boson Hamiltonian to produce the model Hamiltonian
(\ref{EQ7}). For this very reason only the lower part of the figure around the
critical $\alpha_{iso}$ belongs to the validity region of the approximation.
Finally, an analog of the relative populations of the two modes is the
semiclassical imbalance $\zeta_b$ (defined only for the coherent states). It can be
cast as
\be
\zeta_b = -\frac{\gamma}{1-\Lambda}=\frac{\chi_{22}-\chi_{11} +
2(E_2-E_1)\Omega/(gN)}{\chi_{11}+\chi_{22}-2\chi_{12}},
\en{EQ21}
see Fig. \ref{FG5}(b).

Finally, let us make some comments on the relative phase $\phi$. Why the phase
$2\phi$ appears in the classical Hamiltonian $\mathcal{H}_\mathrm{cl}$ (\ref{Hcl})
is clear: the bosons tunnel by \textit{pairs}, which is reflected in the splitting
of the even and odd subspaces of the Fock space, with the respective basis states
$|2s,N-2s\rangle$ and $|2s-1,N-2s+1\rangle$ \cite{SKPRL,ParEff}.  Since the state
of the system is always expanded over the states differing by an even number of
bosons, it is impossible to define the phase $\phi$, but only the $2\phi$: $2\phi =
\mathrm{arg}(\langle(b^\d_1)^2b_2^2\rangle$). Hence $2\phi$ and not $\phi$ appears
in the exponent factor in Eq. (\ref{EQ11}): $\exp\{i\phi \zeta/h\}
=\exp\{2i\phi(n_1-n_2)\}$. The splitting of the Fock space into two subspaces also
leads to the double  degeneracy of the coherent states (quasi-degeneracy to be
precise: the terms of order $1/N$ are neglected), since the same approximate
``wave-function'' in the Fock space $\psi(\zeta)$ describes not one but two states,
one of each subspace: $C_{2s}= \langle 2s,N-2s|\psi\rangle$ and $C_{2s-1} =\langle
2s-1,N-2s+1|\psi\rangle$ with the discrete sets $\zeta_1 \in\{ (2s-1)/N-1\}$ and
$\zeta_2\in \{2s/N-1\}$.

The mean-field  approach, in contrast, produces a definite relative phase, see Ref.
\cite{CW}, where two equivalent  order parameters of the nonlinear Gross-Pitaevskii
equation are possible  for the description of the same experiment with the phase
either $\pm\pi/2$, due to the broken superposition principle by the nonlinearity.
However, the full many-body quantum Hamiltonian permits superposition of the
eigenstates of the same energy. The resolution of this seemingly paradoxical
situation is similar to the case of the random phase in  the double-slit experiment
with the Bose-Einstein condensate, see Ref. \cite{QPhase}.  Indeed, since the atoms
are detected one by one coherently from both modes $b_{1,2}$, when the lattice is
released, the atom detections  probe the quantity  $\langle b^\d_1b_2\rangle$
spontaneously projecting, as the detection process proceeds,  on one of the two
possible phases $\phi_b = \pm\pi/2$ of $C_\pi$.

In conclusion, we have shown that the experiment  of Ref. \cite{Pboson} is
describable by the quantum model (\ref{EQ7}) and that there is the quantum phase
transition of the second order between the atom number-squeezed states and the
coherent phase states of the $p$-bosons. The results indicate that in the recent
experiment \cite{Pboson} a phase transition of the second order was observed, where
the isotropic experimental state observed for the symmetric point
$\alpha=\mathrm{arccos}\epsilon$ (and hence, for $\gamma = 0$) must be the coherent
$C_\pi$ state of the relative phase $2\phi=\pi$.

\acknowledgments   This work was supported by  the FAPESP  and CNPq of Brazil.

\end{document}